\documentclass[preprint2]{aastex}
\usepackage{epsfig}
\usepackage{natbib}
\usepackage{lscape}
\usepackage{subfigure}
\usepackage{amssymb,amsmath}

\DeclareMathVersion{bold}

\begin{document}

\shorttitle{``Extragalactic'' Bursts Show Terrestrial Origins}
\shortauthors{Burke-Spolaor et al.}

\title{Radio Bursts with Extragalactic Spectral Characteristics Show Terrestrial Origins}

\author{S. Burke-Spolaor\altaffilmark{1} and Matthew Bailes}
\affil{Ctr. for Astrophysics and Supercomputing, Swinburne University of Technology, Mail H39, PO Box 218, Hawthorn VIC 3122, Australia}
\email{sburke@astro.swin.edu.au}

\author{Ronald Ekers}
\affil{CSIRO Australia Telescope National Facility, PO Box 76, Epping NSW 1710, Australia}

\author{Jean-Pierre Macquart}
\affil{ICRAR/Curtin Institute of Radio Astronomy, GPO Box U1987, Perth WA 6845, Australia}

\and

\author{Fronefield Crawford III}
\affil{Dept. of Physics and Astronomy, Franklin and Marshall College, Lancaster PA 17604}

\altaffiltext{1}{Co-affiliated with CSIRO Australia Telescope National Facility, PO Box 76, Epping NSW 1710, Australia}

\begin{abstract}
  Three years ago, the report of a solitary radio burst was thought to be the
  first discovery of a rare, impulsive event of unknown extragalactic origin
  (Lorimer et al. 2007). The extragalactic interpretation was based on the
  swept-frequency nature of the event, which followed the dispersive delay
  expected from an extragalactic pulse.  We report here on the detection of 16
  pulses, the bulk of which exhibit a frequency sweep with a shape and
  magnitude resembling the Lorimer Burst. These new events were detected in a
  sidelobe of the Parkes Telescope and are of clearly terrestrial origin, with
  properties unlike any known sources of terrestrial broad-band radio
  emission.  The new detections cast doubt on the extragalactic interpretation
  of the original burst, and call for further sophistication in radio-pulse
  survey techniques to identify the origin of the anomalous terrestrial
  signals and definitively distinguish future extragalactic pulse detections
  from local signals.  The ambiguous origin of these seemingly dispersed,
  swept-frequency signals suggest that radio-pulse searches using multiple
  detectors will be the only experiments able to provide definitive
  information about the origin of new swept-frequency radio burst detections.
\end{abstract}

\maketitle


\section{Introduction}
Extragalactic radio phenomena emitting transient radio bursts have been
theorized to arise from a myriad of violent cosmic events, including
coalescing systems of relativistic massive objects
\citep[][]{relativisticcoalesce,hansenlyutikov01}, the evaporation of
primordial black holes \citep{rees77}, and supernova events
\citep[e.\,g.][]{colgatenoerdlinger71}.  The frequency-dependent dispersive
delay (at $\delta t\propto\nu^{-2}$) characteristic of a radio pulse that has
propagated through the cold plasmas of the interstellar and intergalactic
media encodes the source's distance and the line-of-sight free electron
density, furthermore enabling extragalactic radio pulses to be used as robust
cosmological probes of the ionized content of the intergalactic medium.
However, such pulses have proven difficult to detect. There has thus far been
a single claim for the detection of an extragalactic pulse: the discovery that
has come to be called the ``Lorimer Burst'' \citep[hereafter LB,][]{lorimer}.
The strongest evidence of its extragalactic origin was a large delay at
$\delta t\propto\nu^{-2}$ (suggesting a dispersed signal, with a ``dispersion
measure'', \mbox{DM $=375$\,pc\,cm$^{-3}$}, where the DM quantifies the
integral electron density along the line of sight to the emitter), indicating
a distance of well outside of the Galaxy when the electron content of our
Galaxy is accounted for \citep{ne2001}. It furthermore appeared to exhibit the
frequency-dependent Komolgorov scattering that is expected from signals
propagating in the interstellar medium.  Based on its detection in three of
the thirteen beams of the Parkes multibeam receiver at the expected relative
levels for a point source, the burst appeared to be coming from the sky.

Since its discovery, some doubt has emerged over the extragalactic origin of
the LB. Its extreme intensity ($\sim$100 times the detection threshold)
implies that searches of similar sensitivity should detect fainter events, if
such bursts are isotropically distributed throughout the Universe and/or have
an underlying intensity distribution typical of astrophysical phenomena. The
failure of further searches to find similar events indicates that either this
is not true, that the astrophysical process that caused the Lorimer Burst is
extremely rare ($<2.8\times10^{-5} {\rm hr} ^ {-1} {\rm deg} ^{-2}$ for fluxes
$S>300$\,mJy, based on the data from LB, \citealt{deneva}, \citealt{keane},
\citealt{sbsmb}), or the burst's properties were misinterpreted.

We announce the detection of 16 pulses that exhibit a frequency-swept signal
with similar characteristics to the Lorimer Burst.  However, these new events
are clearly of terrestrial origin with properties unlike any known sources of
broad-band radio emission. We detail the properties of the new detections and
scrutinize them in comparison to the LB.  We also discuss methods that can be
used to distinguish an astrophysical or terrestrial origin for future
detections of frequency-swept pulses.

\section{Data sets and search}\label{sec:data}
Prompted by the discovery of the Lorimer Burst, we searched 1078 hours of data
for signals exhibiting a $\delta t\propto\nu^{-2}$ cold plasma dispersion
delay. The data were archival pulsar surveys taken over the years 1998 to 2003
with the 20cm multibeam receiver installed at Parkes Telescope.  Four surveys
were searched. For two \citep{SM,BJ} we followed the search and inspection
process described by \citet{sbsmb}, differing only in interference mitigation
filters; we inspected a candidate if it either a) was detected at a
signal-to-noise ratio of $>6$ in less than 9/13 of the receivers, b) had a
signal-to-noise ratio of $\geq20$, or c) showed a DM of higher than
250\,pc\,cm$^{-3}$. As these were surveys of high Galactic latitude ($|b| >
5^\circ$), we searched DMs up to 600\,pc\,cm$^{-3}$. For the remaining two
surveys \citep[][their observing parameters are detailed in their Sec.  2 and
Table 1, respectively]{froneydata1,froneydata2}, we used methods described by
\citep{froneysearch1,froneysearch2} applying no interference excision based on
multiple-beam detections. These were generally surveys of lower galactic
latitude, therefore we searched DMs up to 1000 \citep{froneydata1}, 2000 (for
target AX J1826.1Ð1300 of the Roberts et al. survey) and 2500\,pc\,cm$^{-3}$
(for the other targets of Roberts et al.). Candidates from these surveys were
inspected one beam at a time by eye to identify dispersed pulses.

\section{New discoveries and their properties}\label{sec:properties}
Our search revealed 16 pulses with two striking features that distinguish them
from all others in the data: an apparent $\delta t\propto\nu^{-2}$ delay of a
magnitude implying an extragalactic origin in the telescope's pointing
direction, and a simultaneous occurrence in all 13 telescope receivers at
relative intensities of less than a factor of four (Fig. 1). When a dispersive
delay is fit to each detection, the values cluster about a net band delay
$\Delta t = 360$\,ms, indicating a close connection with the LB at $\Delta t =
355$\,ms (Fig. 2).  The LB's reported sky position was below the horizon for
several detections, therefore the pulses could not have come from the same
extragalactic source. Below, we give evidence that the 16 signals have a
terrestrial origin.

The 29$^{'}$ separation between each receiver's beam position and $>20$\,dB
attenuation beyond 30$^{'}$ from each beam center render it impossible for an
on-axis, pointlike signal to appear in more than 3 beams at similar intensity
\citep{multibeam,feedlegs}. Out detections were therefore made through a
sidelobe of the Parkes antenna, and based on the consistency of signals in the
beams, in each case the emitter was positioned $\gg$5 degrees from the
telescope's pointing direction. Consequently: 1) we have a
minimum-senisitivity field of view of $\sim$20000\,deg$^{2}$, a detection rate
$2.3\times10^{-7}\,{\rm deg}^{-2}\,{\rm hr}^{-1}$ (0.1 day$^{-1}$ with our
observing system), and poor accuracy for emitter localisation, 2) the pulses
are subject to frequency-dependent dropouts and scattering from multi-path
propagation---visible in Fig. 1(b-d) and explored below---and 3) the
source(s)' intrinsic flux density, had we pointed directly at it, is a factor
of 2500-850000 greater than the detected value ($0.8 < S_{intrinsic} <
272$\,kJy for the brightest detection, $0.1 < S_{intrinsic} < 34$\,kJy for the
faintest; see Table 1).

We conclude a terrestrial origin for these bursts based on their extreme
brightness and two other features. First, some exhibit deviations from a model
dispersive delay: e.g. the sharp kink at 1465 MHz in detection 06 (Fig. 1c),
and subtler deviations in other detections (see the $\chi^2$ listing in Table
1). Despite a trend mimicking that expected from dispersion, such deviations
decisively distinguish the pulses' frequency-dependence from a delay induced
by interstellar propagation.  Hereafter we distinguish these detections with
the name ``Perytons,'' representing the non-dispersive, highly swept,
terrestrial signals exhibited by the pulses.\footnote{The name is chosen from
  mythology to be unassociated with an exact physical phenomenon, due to the
  ambiguous origin of the detections; Perytons are winged elk that cast the
  shadow of a man.}

Second, the temporal distributions of the signals strongly imply a terrestrial
origin (Table 1). All detections occurred during daylight, primarily
mid-morning. Eleven appeared in one 4.4-minute observation followed by another
0.5 hours later (we regard these as non-independent), while the remaining
events occurred in isolation. Four of the five independent detections appeared
in a 3-week period in late June/early July spread across the years 1998 to
2003, coinciding with the peak of Australian mid-winter. Their time
distribution appears to follow a non-random both annual and daily cycle; we
tested the probability that 4/5 events would occur in June/July given an
underlying random distribution by running a Monte-Carlo simulation based on
the monthly hours observed. This test resulted in a confidence of P = 0.997231
of a non-random annual distribution. A similar test based on the time of day
distribution of observations and detections gave a probability of P = 0.999046
that 4/5 events would occur in the UT range 0--3. These cycles are strongly
suggestive of either a climate/weather-related effect, or a man-made origin
for the emission.

\section{Discussion}
\subsection{Signal origins}\label{sec:origins}
It is unprecedented for non-astrophysical emission to exhibit such drastic
frequency-dependent delays in the 1 GHz band. Given the daytime occurrence of
the Perytons, we first explored the possibility of the signals as man-made.
The continuous emission across the legally protected 1400-1427 MHz band
suggest that the signal is not intentionally transmitted; additionally, the
lack of regular periodicity, broad pulse widths (30-50 ms), and broad-band
emission preclude a radar origin. Man-made emission that is unintentionally
transmitted arises often from on-site electronic hardware failure. This does
not appear to be the source of this emission, however, based primarily on the
amplitude modulation seen in all the detections. These amplitude-modulated
temporal or frequency structures show conformity across the spectrum in all 13
beams for each burst. Assuming the modulation is attributable to multi-path
propagation effects (that these detections will necessarily show, as noted in
\S\ref{sec:properties} above), the incoming wavefront must not decorrelate
over the physical size of the telescope's feed horns (1 meter), to allow the
similarity of the modulation structures in all 13 beams. The diffractive scale
($s_0$), therefore, must likewise be $\geq1$\,m. Taking the characteristic
bandwidth of the modulation structure in all pulses to be $\Delta
f\sim10$\,MHz), and based on the center frequency $f = 1374$\,MHz, we place a
lower limit on the distance to the object(s) and scatterer(s) of
$2\pi(s_0f)2/c\Delta f > 4$\,km. This suggests that we have detected emission
from the horizon---well off-site from the telescope grounds---and provides the
strongest argument against on-site hardware failure as the source. The
quasi-annual cycle and the spectral complexity with a 6-year persistence of
the signal also argue against local hardware failures as the emission's
origin.

It is likewise possible that the Perytons were caused by a natural terrestrial
source. We explored this possibility, considering processes that can produce
non-dispersive, swept emission with sweep rates of $\sim$1\,GHz/s. The
emission requires a process of finite-bandwidth ($\Delta f < 25$\,MHz) signal
to progressively change in its center frequency, for instance cyclotron
emission in a time-varying magnetic field or the progressive incitement of
plasma oscillations in regions of differing plasma density. The latter of
these processes occurs in type III solar bursts \citep[e.\,g.][]{solarbursts},
and it may be possible for such a process to occur in Earth's atmosphere with
lightning or transient luminous event phenomena \citep[e.g.][]{su,sentmann},
which can reach the necessary ionization levels for Langmuir-wave
electromagnetic emission to occur in our observing band.


\subsection{A closer look at the Lorimer Burst}
Whether the LB has the same origin as our detections remains inconclusive,
despite the serious doubt these signals add to its extragalactic
interpretation. Several qualitative differences between the LB and the
Perytons warrant examination. The LB was clearly detected in only 3 of the 13
beams (6, 7, and D). It is marginally detectable in other beams, most
prominently in beam C, and apparent in a stacked time series of the remaining
beams. Based on observations of the inner sidelobe pattern of the multibeam
obtained by L. Staveley-Smith (private communication), we find a position of
the LB (consistent with both the relative detected flux levels in beams 6, 7,
C, D, and the non-detections in the other beams) at RA $19.44\pm0.08$, Dec
$-75.17\pm0.08$. That is, the relative signal levels of the LB conform to
those expected from a boresight signal, in agreement with the same conclusion
of \citet{lorimer}. Therefore, if the LB is a Peryton, it appears to be the
only detection for which the telescope was pointed directly at the emitter.
Consequently, although the LB did not exhibit the same deep spectral signal
modulation as the Perytons, these differences are well accounted for by the
multi-path effects which we interpreted to arise from the horizon-based
sidelobe detections of the Perytons (\S\ref{sec:origins}).

We note that because the beam and sidelobe shapes of the multibeam scale with
frequency, the offset of the burst from the center of beam 6 will induce a
spectral steepening of the source, causing the intrinsic spectrum to be
flatter than that originally reported by \citet{lorimer} by $\alpha_{\rm
  intrinsic} = \alpha_{\rm observed}-\alpha_{\rm induced}$ (where
$S\propto\nu^{\alpha}$). The Staveley-Smith measurements were made at two
frequencies, allowing quantification of this effect. Within the error of our
positional measurements, the induced index is $\alpha_{\rm induced} =
-1^{+0.9}_{-2}$. Therefore, at our estimate of position, the intrinsic
spectral index is $\alpha_{\rm intrinsic} = -2.5\,{\rm to}\,-0.6$, where we
have measured the observed spectrum in beam 6 to follow $\alpha_{\rm
  obs}=-2.6$.


If the LB was caused by a sky-based object (e.\,g. an aircraft or a natural,
propagating swept-frequency phenomenon), we might expect to detect some
movement of the LB across the field of the multibeam. We can limit the
movement by noting that because the signal saturated beam 6 for the entirety
of its sweep, it is clear that the emitter did not travel sufficiently far to
cross a null in the sidelobes of beam 6. Therefore, we limit any movement of
the emitter to $<35$\,arcminutes, corresponding to a distance of
$d=0.015\,h/{\rm sin(\theta)}$, where $h$ is the emitter's altitude and
$\theta$ is the angle between the telescope's line of sight and the LB's
velocity vector. At 12\,km (a typical aircraft/cloud height), this
corresponds to a distance and velocity of 180\,m and $500$\,m/s, respectively,
if the LB is moving perpendicularly to the line of sight. This does not put
rigorous limits on aircraft movement, however does place bounds on propagating
atmospheric phenomena which could give rise to this emission. An analysis
which places a more stringent limit on (or provides a measurement of) movement
of the LB would be possible with a more sophisticated electromagnetic beam
model.

One major point of discrepancy remains between the LB and the Perytons: that
their widths disagree by a factor of $\sim$2 (which cannot be accounted for by
multi-path scatter broadening), and we do not observe a frequency-dependent
pulse width evolution in the Perytons. However, we are hindered in measuring
frequency-dependent evolution in the Perytons because of their modulated
signal. While the difference in pulse width lends weak remaining support for a
divide between the LB and the Perytons, with our current measurements it
cannot be ruled out that there may be an underlying pulse width distribution,
and/or a dependence on width with an event's intrinsic flux; likewise we
cannot state whether frequency dependence of pulse width is intrinsic to the
phenomenon causing all the events.

\section{Conclusions: Implications for current and future transient experiments}
Regardless of the physical origin of these pulses or the LB, the results of
this study illustrate the limitations of single-dish radio burst detection
experiments to provide conclusive evidence for the origins of one-off bursts;
dispersive delays can provide the only evidence for an astrophysical nature on
single-detector detections, while multi-detector (i.\,e. array) experiments
can provide wavefront measurement and localization for localized pulses, or
can assure non-correlation of local signals between widely-spaced array
elements. Array experiments are necessary to provide a conclusive origin for
further detections of the class of pulses presented here, as well as for any
future experiments that aim to detect and use extragalactic pulses in
scientific studies. Two such experiments are currently underway at the Giant
Metre Wave and Very Long Baseline Array telescopes, and will be possible with
the Square Kilometre Array and its pathfinder experiments. Single-dish
measurements can improve our understanding of these events by providing
polarisation measurements, and further detections for world, temporal, and
delay-distribution statistics.

If our detections do originate from a natural terrestrial process with
intrinsic flux $\gg$100\,Jy, detections are expected at other observatories
with similar capabilities to Parkes and ongoing pulsar and transient
observations (e.g. Arecibo Observatory, Green Bank and Effelsberg Telescopes).
It is an undeniable curiosity that no Perytons have yet been identified at
radio observatories where such searches have been performed
\citep[e.\,g.][]{shaun,nice,deneva}. However, it is possible that they have
not occurred during observations (one should occur per $\sim$215\,h of data on
a telescope with similar system temperature and sidelobe suppression levels to
Parkes, if the event rate is the same at other sites), that interference
rejection algorithms based on the appearance of signals in multiple beams have
removed the signals, that the searches were at frequencies where Perytons do
not emit, and/or that the experiments were not sensitive to pulses of
30-50\,ms width. As multi-detector coincidence filters will strongly preclude
the detection of Perytons, the strength of such rejection filters to limit
local interference of any form is clear; however it is likewise clear that the
filters would need to be relaxed to maximize an observatory's capability to
detect and identify the origin of these remarkable signals. Additionally, we
believe that it was the human inspection of the spectrogram data and
multi-beam time series for each candidate (described in \S\ref{sec:data}) that
were the main contributors to the first recognition of the peculiar nature of
the Perytons---the simultaneous occurrence of clear, seemingly dispersed
emission and multiple-beam detection---and that encouraged us to scrutinize
these events more closely instead of disregarding them as spurious detections.

\section{Acknowledgements}
The authors gratefully acknowledge valuable discussions with M. Kesteven, C.
Wilson, P. Burford, J. Reynolds, D. Melrose, P. Kalberla, and P. Edwards about
various aspects of this research. Financial support was granted by the
Research Corporation and the Mount Cuba Astronomical Foundation (FC), and the
Australian Research Council (MB).

\clearpage

\begin{figure*}
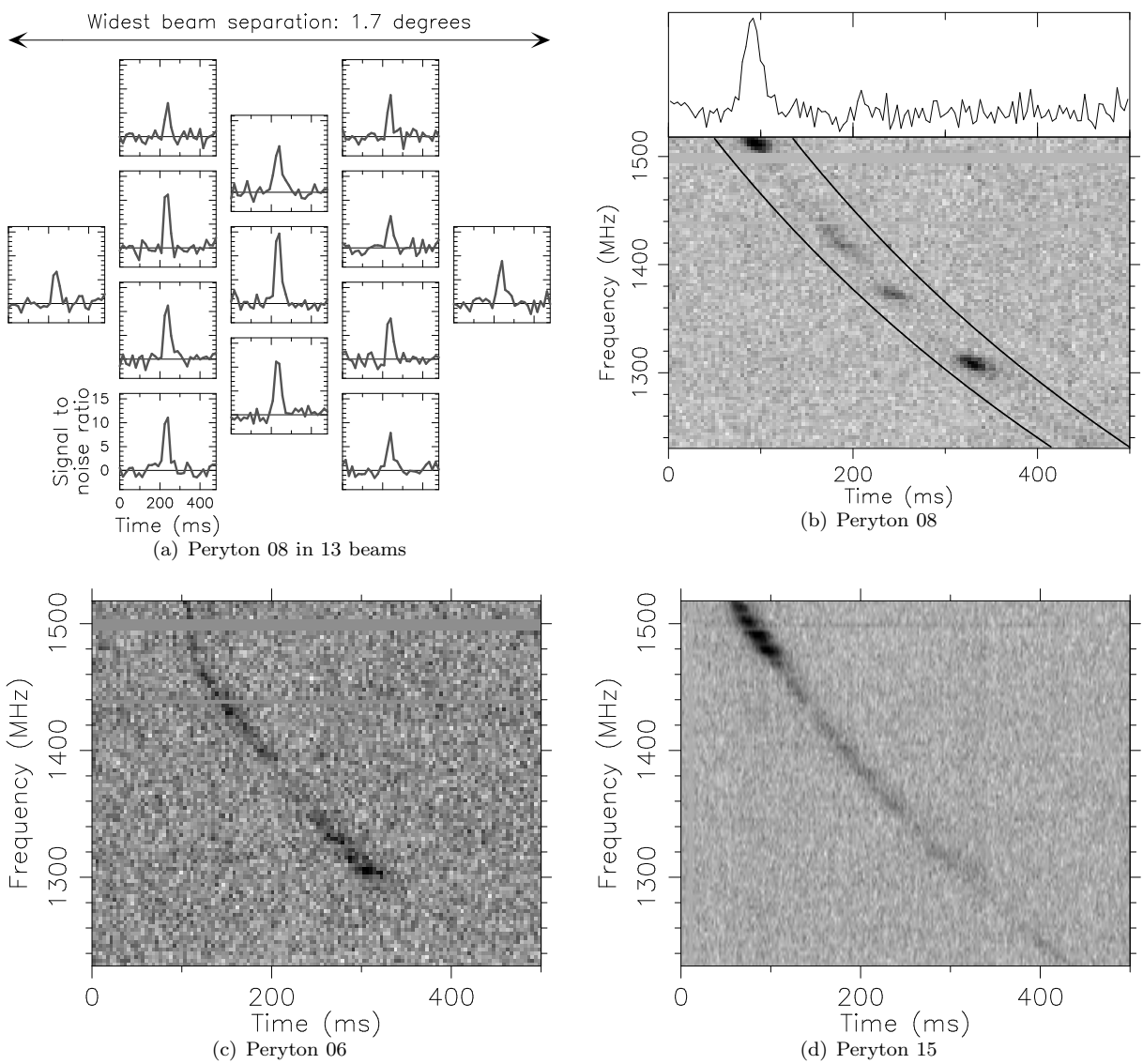

\epsscale{1.0}
\centering
\subfigure[Peryton 08 in 13 beams]
{
  \includegraphics[height=0.47\textwidth,angle=270]{1a.ps}
}\quad
\subfigure[Peryton 08]
{
  \includegraphics[height=0.47\textwidth,angle=270]{1b.ps}
}\\
\subfigure[Peryton 06]
{
  \includegraphics[height=0.47\textwidth,angle=270]{1c.ps}
}\quad
\subfigure[Peryton 15]
{
  \includegraphics[height=0.47\textwidth,angle=270]{1d.ps}
}
\caption{Spectrograms and time series for several detections. {\bf(b,c,d)}
  Data from the 13 beams have been summed to enhance the signal. Frequency
  channels with known interference have been blanked. {\bf(a)}, De-dispersed
  time series showing Peryton 08 in the 13-beam multibeam receiver as the
  beams are distributed on the sky. {\bf(b)} De-dispersed time series and
  spectrogram of Peryton 08. The black lines trace the best-fit dispersive
  delay for this detection. {\bf (c,d)} Spectrograms of Peryton 06 and 15,
  respectively.}
\end{figure*}

\clearpage

\begin{figure*}
\centering
\includegraphics[height=0.5\textwidth,angle=270]{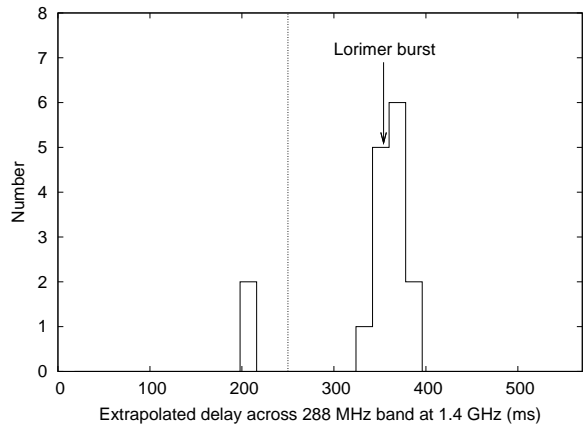}
\caption{Distribution of fitted dispersive delays. The x-axis shows the total
time to cross the observing band, calculated from each pulse as a best-fit
dispersive sweep. An arrow indicates the delay of the Lorimer Burst. The
dotted line indicates the delay below which we had uneven search criteria (see
\S\ref{sec:data}).}
\end{figure*}

\clearpage

\begin{table*}
  \centering
  {\footnotesize
  \begin{tabular}{cccccccccccccc}
\hline
 \hline
{\bf Peryton}& {\bf UT}                  & $\theta_z$& $\theta_a$& {\bf $\Delta t$}&          & $w$ & $\bf{S}_{\rm {\bf det}}$ \\
{\bf ID \#}      & Y-M-D-h:m:s         &      (deg)    & (deg)          & (ms)     & $\chi^2$ & (ms) & (mJy) \\
 \hline
 \hline
01& 98-06-23-02:03:44.91 & 33.341 & 136.657 & 381.9 & 7.0 & 35.2 & 90 \\  
02& 98-06-23-02:04:06.75 & 33.288 & 136.692 & 352.6 & 2.5 & 46.9 & 90 \\  
03& 98-06-23-02:04:28.84 & 33.235 & 136.728 & 362.0 & 2.1 & 31.2 & 90 \\  
04& 98-06-23-02:04:36.84 & 33.216 & 136.740 & 356.4 & 2.8 & 35.2 & 100 \\ 
05& 98-06-23-02:05:17.77 & 33.118 & 136.807 & 354.4 & 1.5 & 35.2 & 70 \\  
06& 98-06-23-02:05:39.50 & 33.066 & 136.843 & 343.1 & 8.0 & 31.2 & 70 \\  
07& 98-06-23-02:06:01.81 & 33.013 & 136.879 & 363.0 & 2.4 & 39.1 & 80 \\  
07a&98-06-23-02:06:24.13 & 32.960 & 136.916 & 363.9 & --- & 32.1 & 40 \\  
08& 98-06-23-02:06:31.89 & 32.941 & 136.930 & 369.6 & 4.6 & 39.1 & 100 \\ 
09& 98-06-23-02:07:27.70 & 32.808 & 137.023 & 328.9 & --- & 43.0 & 60 \\  
10& 98-06-23-02:07:49.78 & 32.755 & 137.061 & 349.7 &4.7\tablenotemark{a} & 31.2 & 60 \\
11& 98-06-23-02:34:53.63 & 29.738 & 136.640 & 360.1 &1.9\tablenotemark{a} & 46.9 & 320 \\ 
12& 98-06-25-05:26:49.13 & 25.445 & 141.515 & 363.9 & 0.8 & 39.1 & 110 \\ 
13& 02-03-01-01:25:38.88 & 34.519 & 320.875 & 207.0 & 1.3 & 31.3 & 110 \\ 
14& 02-06-30-02:10:29.38 & 28.465 & 189.173 & 203.2 & 2.4 & 39.1 & 240 \\ 
15& 03-07-02-00:09:23.96 & 44.092 & 000.631 & 378.1 & 4.9 & 39.1 & 220 \\ 
 \hline
LB  & 01-07-24-19:50:01.63 & 42.419 & 183.315 & 354.5 & 1.6 &15.6 & 30000\\
 \hline
 \hline
 \end{tabular}
 \caption{{\small Columns: (1) Chronological ID; Peryton 07a was discovered after summing the
13-beam data, therefore has a non-standard index. The Lorimer Burst
is given for reference; (2) U.T. arrival time at 1516.5\,MHz; (3,4) Telescope
zenith and azimuth angle, respectively, at the time of detection; (5)
Extrapolated best-fit DM delay across the band; (6) Reduced-$\chi^2$ for a
$\delta t\propto\nu^{2}$ fit to the burst, based on the event's time of
arrival in 48\,MHz sub-bands (if ${\rm S/N_{band}}>5$), timed against an
analytic model of the event's de-dispersed profile at the best-fit DM. Events
07a and 09 had no sub-bands of ${\rm S/N_{band}}>5$; (7) Event width at half
maximum after de-dispersing at the best-fit quadratic delay; (8) Detected
single-beam peak flux, defined $S_{\rm det}={\rm S/N}\cdot
T_{sys}/(G\sqrt{N_{\rm pol}wB})$, where S/N is the detection's signal-to-noise
ratio. The Perytons' intrinsic flux is much greater than reported here (see
main text).}}}
\tablenotetext{a}{This value would decrease without the presence of strong interference in the observation.} 
\end{table*}


\end{document}